\documentclass[12pt]{article}

\usepackage{hyperref}
\usepackage{breakurl}
\begin{document}

\centerline{\bf\large Astronomy Career Profiles from the AAS Newsletter Archives}

\begin{center}
{Travis Metcalfe, Leila Belkora, Liam McDaid, Blake Bullock, Christine Pulliam, \\ 
Peter Williams, Joshua Roth, Barb Whitney, Knut Olsen, Andy Howell, Luke Keller}
\end{center}

\begin{abstract}
This is a collection of articles that were originally published in the 
Newsletter of the American Astronomical Society (AAS) between May 2008 and 
September 2011 by the Committee on Employment. Authors representing a wide 
range of career paths tell their stories and provide insight and advice 
that is relevant to success in various job sectors. Although all of these 
articles are available individually from the AAS archives, we are posting 
the complete collection here to make them more accessible as a resource 
for the astronomy community. The collection includes the following 
articles: (1) Changing Priorities: the Hard Money Wild Card, (2) Beyond 
Ivory Towers, (3) Astronomers Working in Public Outreach, (4) 
Bush-Whacking a Career Trail, (5) Science Communication as a Press 
Officer, (6) Jobs in Industry, (7) Back to School: A Ph.D.\ Enters the 
Classroom, (8) Working at a Soft-Money Institute, (9) Balancing Research 
and Service at NOAO, (10) Succeeding in a Large Research Collaboration, 
and (11) Preparing for the College Teaching Job Market. The final published
versions of the articles can be found at \url{http://aas.org/career/}
\end{abstract}

\newpage

\centerline{\bf\large Changing Priorities: the Hard Money Wild Card}
\vskip 6pt
\centerline{Travis Metcalfe (\href{mailto:travis@ucar.edu}{travis@ucar.edu})}
\vskip 4pt

If you are an astronomer who doesn't work at a university, the chances are 
good that you work at one of the many federally funded research facilities 
or observatories (e.g. STScI, NOAO). There are several types of positions 
at such institutions, including some supported by grants (soft money) and 
others supported by base funding from the sponsor (hard money). Some hard 
money positions can even be on a tenure track, with the usual disclaimer 
about being ``contingent on the availability of funding''. These jobs 
generally involve some combination of research and service in support of 
the mission of the organization. Such positions are ideal for scientists 
who want to spend most of their time doing research rather than teaching 
-- the only catch is that your research must be relevant to the strategic 
goals of the institution.

My first experience working in a federally funded research laboratory came 
during the summer before I finished my Ph.D. One of the external members 
of my thesis committee was a hard money scientist at the High Altitude 
Observatory (HAO) in Boulder, Colorado -- part of the National Center for 
Atmospheric Research (NCAR), which is sponsored directly by the National 
Science Foundation. Growing up professionally in an academic environment, 
I was surrounded by scientists who decided that being a university 
professor was the best career choice. This was my first exposure to an 
institution filled with people who had made a different evaluation. It was 
a powerful experience that really resonated with my vision of the ideal 
job, and by the end of the summer I was convinced that a hard money career 
path was right for me.

After finishing my thesis, I spent several years as a postdoc before 
landing an NSF fellowship that brought me back to HAO. About a year later, 
I was hired as a tenure track hard money scientist. As the name suggests, 
the primary research focus at NCAR is the Earth's atmosphere -- but 
because the Sun is responsible for the energy input at the top of the 
atmosphere and for the particle flux underlying disruptive ``space 
weather'' events, HAO is the NCAR laboratory devoted to solar physics. My 
job was to maintain a connection between this group of several dozen solar 
physicists and the wider astrophysics community -- using stars to provide 
a broader context for our understanding of the Sun, and ensuring that 
stellar research could benefit from the laboratory's detailed knowledge of 
our local star.

Life as a hard money scientist was good. In addition to having access to a 
12-month salary without the requirement of writing grant proposals, each 
scientist was allocated a modest annual travel stipend while internal 
funds also paid for journal page charges and even helped bring in 
scientific visitors. In return, we worked on large-scale and long-term 
projects that were not amenable to funding through standard three-year 
grants, often with a focus on serving the scientific community with new 
modeling capabilities or public data. The primary disadvantage of being an 
astronomer in a solar physics laboratory was the difficulty of finding 
students and postdocs. Unlike a university environment where students are 
the lifeblood, only a few students could be supported by internal 
fellowships at HAO. Postdocs were also hard to find, since most of the 
fellowship applicants were interested in solar physics, not astronomy. 
Consequently, like many hard money scientists I still wrote grant 
proposals to help recruit students and postdocs, and to provide part of my 
salary.

Perhaps the greatest source of anxiety for a hard money scientist is the 
annual drama of the federal budget cycle. Flat budgets at the federal 
level generally translate into a flat budget for the NSF and all of its 
programs. As everyone knows, a flat budget in the face of rising operating 
costs really means a cut. When the budgets do get increased for inflation, 
salary levels within each laboratory are supposed to be adjusted according 
to merit -- but in reality the extra funds either disappear entirely to 
offset a previous budget shortfall, or they are distributed evenly among 
the staff to compensate for the years without a cost of living adjustment. 
The leadership in Washington certainly recognizes the importance of 
scientific research as an engine of economic growth and innovation (read 
the America COMPETES Act), but these lofty pronouncements rarely seem to 
be reflected in national budget priorities.

Nobody needs to be reminded of the chaos surrounding the most recent 
federal budget cycles. A series of short-term ``continuing resolutions'' 
to fund the government at 2010 levels ultimately led to a budget for 2011 
that finally passed more than halfway through the fiscal year. By this 
time the NSF and its programs were already preparing budget scenarios for 
2012, and the partisan rancor in Washington made it clear that difficult 
decisions were unavoidable. It was in this atmosphere that HAO concluded 
it could no longer support stellar research, and I was given 12 months 
notice that my position would be eliminated. Despite outstanding annual 
performance reviews and wide ranging contributions to programs across the 
organization, changing priorities motivated by federal budget cuts ended 
my career as a hard money scientist.

Fortunately there are other ways to survive as a research scientist. After 
my final year on hard money, I hope to continue working at NCAR for 
another year or so on soft money. In the longer term, I will probably need 
to seek an environment with lower overhead expenses to continue funding 
myself on grants. As a graduate student I formed a non-profit organization 
dedicated to scientific research and public education, thinking that it 
could always be my backup plan in case a hard money position didn't 
materialize. This unexpected career transition may be just the impetus I 
needed to build on this foundation, and hopefully make a soft landing on 
soft money. Wish me luck.

\vskip 6pt
{\it AAS Newsletter, 160 (Sep/Oct 2011)}

\newpage
 
\centerline{\bf\large Beyond Ivory Towers}
\vskip 6pt
\centerline{Leila Belkora (\href{mailto:Belkora@cox.net}{Belkora@cox.net})}
\vskip 4pt

My career took a non-traditional turn as I finished my Ph.D.\ in 1995. I 
had always been interested in science writing and I was offered a great 
job as an editor in the Office of Public Affairs at Fermilab in Illinois. 
I figured I would try it out for a year, and would still be able to look 
for a postdoc position if it didn't work out.

My responsibilities there mainly involved writing about particle physics 
and astro-particle physics for the public. I was already a good writer but 
I learned a lot from my boss about writing for the public. (Who knew that 
``beam'' is jargon, and that neutrinos can be scary to the uninitiated?) I 
had a blast meeting the lab's researchers and technicians and writing 
about them and their work for FermiNews. I also talked to reporters, 
showed visitors around and helped write the lab's annual report. I missed 
being involved in research, but I enjoyed the more social work environment 
and the writing. My former peers in astronomy were not nearly as negative 
about my defection as I had feared; that was reassuring too.

After a few years I took a job as science and engineering publicist in the 
public affairs office of the University of Illinois at Chicago. My main 
activities were similar to those I carried out at Fermilab, but in my 
``spare'' time I taught physics in the physics department and a history of 
science class in the Honors College. In retrospect, I was not 
well-prepared to teach the large introductory physics classes and it was a 
rather punishing experience, but I loved the small class in the Honors 
College.

I might have continued on this path for years, but around this time my 
husband, a particle physicist, got a short-term position at Cornell. I 
moved with him to Ithaca and decided to write a book during what I thought 
would be a short leave of absence from my UIC job.

At this point I was still contributing a substantial amount to the family 
budget because I was lucky enough to have a large advance for my book (a 
history of astronomy). That soon changed. It took me longer than 
anticipated to finish the book (it finally appeared in 2002), and with the 
extension of my husband's position and the anticipated birth of our 
daughter (2001), I decided to work from home. My contribution to the 
family income has been {\it very modest} ever since, though I have always 
worked and had interesting jobs.

During our time in Ithaca I not only finished the book, but also wrote 
articles for popular astronomy magazines and the Cornell Engineering 
magazine. I taught writing and oral presentation skills to engineers for 
one semester, filling in for a faculty member on medical leave.

In 2002 my husband left academia for industry and we moved to southern 
California, where we are now. I tried to get a position teaching writing 
to engineers, as I had really enjoyed that at Cornell. Instead, I ended up 
grading papers for the online astronomy program at Swinburne University of 
Technology in Australia, which turned out to be equally satisfactory. (My 
title there is ``Project Supervisor''.) Eventually I was also offered a 
job that I hadn't even known existed: developing an online physics course 
through UC-Irvine Extension. My title there is ``Subject Matter Expert.''

I think the quality of life of our family is enhanced by the fact that I 
work from home, and I feel privileged to be able to do so. After taking my 
daughter to school I have about six hours to work. Sometimes I work that 
whole time (taking an occasional break to throw a load of laundry in the 
machine); sometimes I deal with emergencies (taking the car to the repair 
shop seems to occupy a lot of my time lately); and sometimes, I admit, I 
do something fun like visit a new museum exhibit (though this doesn't 
happen a lot). I pick up my daughter from school around 2:30, supervise 
homework, and take her to after-school activities. When my current 
contract work with UC-Irvine is finished, I hope to resume freelance 
science writing (assuming our family circumstances allow it).
 
\vskip 6pt
{\it AAS Newsletter, 156 (Jan/Feb 2011)}

\newpage

\centerline{\bf\large Astronomers Working in Public Outreach}
\vskip 6pt
\centerline{Liam McDaid (\href{mailto:McDaidL@scc.losrios.edu}{McDaidL@scc.losrios.edu})}
\vskip 4pt

Some claims have been made lately that many astronomy students are sold 
sunshine and rainbows when it comes to a career in this field. Some even 
claim that our production of Ph.D.'s is a ``pyramid scheme''. When 
examined, however, the record is clear. There simply aren't huge numbers 
of unemployed Ph.D.\ astronomers.  Whether they are all working in a job 
related to astronomy is another -- respectably heated -- matter. It still 
seems that choosing a career in astronomy always involves big compromises.

A recent development has been the growth of E/PO (Education \& Public 
Outreach) positions. NASA supports some of these, and lately colleges and 
universities are also creating these positions. How these jobs are 
structured varies. The oldest versions had responsibilities involving 
observatory support or running planetariums. They were often connected to 
a science museum or planetarium such as the Fels Planetarium at the 
Franklin Institute. With the plummeting costs of quality imaging and 
spectroscopic equipment, new prospects have arisen. It is possible to do 
original research in an undergraduate class for less than \$25,000 if your 
college already has an observatory (in a dark sky site, of course).

After contemplating how many years I would be poor and frustrated 
otherwise, I sought a job teaching at a community college. Generally, 
although your mileage may vary, an MS is all that is required for such 
positions. In the past, it seemed almost that having a Ph.D.\ counted 
against you in that you were viewed as a ``flight risk'' to a ``real'' 
job. That no longer seems to be true. Given the present economy, this is 
good news, and such positions are rather plentiful. Also, no soft money is 
involved and future funding is pretty stable once you have your foot in 
the door. The bad news is that not every college has the resources to 
support your research...which is something you will be doing in your spare 
time.  Surprisingly, some of my colleagues at community colleges actually 
do research. They have an affiliation with another institution for 
support, however.

My own job is partly (53\%) as an astronomy coordinator. The remainder of 
my time is spent teaching.  As astronomy coordinator, I oversee 
maintenance on our observatory, develop new projects for classes, procure 
needed equipment, function as an astronomy media contact for my area, run 
the observatory for our very popular public viewings, go on location to 
dark sky star parties for our students, create curricula for astronomy 
classes, act as liaison with both local and regional groups on E/PO 
projects and explain the process of science to crowds both enthusiastic 
(astrophiles) and hostile (creationists).

While that last bit may not be to everyone's taste, I feel it is vital to 
get the message of science to as many people as possible -- the same 
people that pay for all the NSF grants, NASA grants, and that infamous 
soft money. As a result, I often find myself dealing with issues that many 
astronomers would rather avoid or wish were not issues. When not answering 
questions about December 2012 for members of the public frightened by the 
apocalypse-industrial complex, I patiently explain why Pluto is no longer 
considered a planet (often a more controversial subject). I do no 
research, although that would be an option for me if I wished. For all the 
trouble my job sometimes pulls me into, it's never boring.

A common theme that may be noticed in talking to people who have 
``untraditional'' astronomy careers is the idea of creating your own job. 
This is often very true. My job was clearly defined on paper, yet many 
things had to be altered or hammered out in the course of actually doing 
the job. There is a certain freedom in this, but for some this prospect 
may seem disturbing. I fear, however, that we are at a point where the 
traditional path is tread by fewer and fewer feet while the majority of 
astronomy graduates walk through different forests. There is no reason to 
think this will change, yet the language is still tilted toward a 
non-university career as being exceptional (unfortunately, said language 
is often salted with vague negative connotations).

The prominence of and often public respect for people in E/PO positions 
has grown as well. Phil Plait, until recently president of the nonprofit 
James Randi Educational Foundation started out in an E/PO position for 
GLAST while working on his first book and blog. He is now working on a TV 
project. Neil deGrasse Tyson also does E/PO as the director of the Hayden 
Planetarium, and has an impressive media profile. For people who have a 
passion for sharing the universe with everyone, this is a great career and 
in the age of 1000 channels and a billion websites it is likely the 
closest thing one can do to follow in the astronomy outreach footsteps of 
the late Carl Sagan.

To find jobs in this growing field, two good sources are the AAS Job 
Register: \url{http://members.aas.org/JobReg/JobRegister.cfm} (usually under 
``Other'') and the ATSC Job Bank: \url{http://www.astc.org/profdev/jobs/jobs.htm}.

\vskip 6pt
{\it AAS Newsletter, 152 (May/Jun 2010)}
 
\newpage
 
\centerline{\bf\large Bush-Whacking a Career Trail}
\vskip 6pt
\centerline{Blake Bullock (\href{mailto:Blake.Bullock@ngc.com}{Blake.Bullock@ngc.com})}
\vskip 4pt

Like many of us, I knew I wanted to work in the field of astronomy from 
the time I could first look up. When asked to draw a picture of what I 
wanted to do when I grew up, I turned in a sheet of paper to my pre-school 
teacher covered from corner to corner in black crayon: space. As I made my 
way through jobs in academics, government, and industry, my enthusiasm to 
learn and contribute never changed, but my notion of where I fit in 
evolved significantly. Here are five things I learned on my personal path.

{\bf People Want to Help:} When I first looked for a job outside the 
standard academic path, it seemed impossible. I had a sneaking feeling 
that websites weren't going to hire me -- a person was going to hire me. I 
asked myself: who do I know who does what I'd like to do? When I found 
someone whose career inspired me, I asked them for fifteen minutes to 
talk. People want to help, but see below.

{\bf People are Busy:} Make it easy for people to help -- ask for 
insights, don't ask for a job. If I only had fifteen minutes, I wasn't 
going to spend it asking questions better suited for Google. As an 
astrophysics undergraduate at Cal Berkeley, my first summer internship was 
at NASA Ames working on the Kepler Mission. The PI needed someone who 
could write and get the word out to the public. Could I do that? Answer: 
yes.

{\bf Be Flexible:} I hadn't expected it, but that summer I learned I loved 
writing about science. The following summer I was hired to write at the 
Lawrence Berkeley National Labs magazine. I kept in touch with the people 
who'd offered me ideas before, and one of my physics professors 
recommended me for the AAAS Mass Media Fellows program 
(\url{www.aaas.org/programs/education/MassMedia/}) where I covered 
the science beat in a newsroom.

{\bf Problem Solvers are in Demand:} After I finished my master's thesis 
in astronomy at Wesleyan University, I wanted to work on a mission from 
the policy side. I applied for the Presidential Management Fellowship 
program (\url{www.pmf.opm.gov}). I was hired as a NASA civil servant. As a 
Presidential Management Fellow, I served at NASA Headquarters, NASA 
Goddard Space Flight Center, and at the Pentagon in the Office of the 
Secretary of Defense, Strategic and Space Programs Directorate. I was most 
enthralled with the opportunity to work on the James Webb Space Telescope 
at NASA, but what shocked me the most was to learn that with my degree in 
astronomy, I was being offered jobs in areas I never expected, such as: 
intelligence, federal law enforcement, and energy.

{\bf A Perfect Fit is Over-rated:} How can I apply for a job at an 
engineering company if I'm trained as a scientist? I did a quick survey of 
the people around me here at Northrop Grumman Aerospace Systems in Redondo 
Beach, CA: the lead for a new mission proposal, our chief scientist, and a 
program's science liaison to NASA. How many of them are specializing in 
the exact area in which they were academically trained? Zero. A strong 
technical background is best for learning how to learn. In industry, where 
technology, processes and the needs of the science community are all 
constantly in flux, on the job training is paramount. The best way I've 
learned what I want to do is by doing. Often times the perfect job for me 
didn't exist until I got there.

As I continue on my career path, I've learned that with every step I take 
in a new direction, I learn something that might be helpful for someone 
just starting out. It's just as important to reach out to others through 
resources like the AAS Non-Academic Astronomer's Network 
(\url{www.aas.org/career/nonacademic.php}). Working on the James Webb Space 
Telescope has opened my eyes to just how many people we need to keep our 
missions going, including inside government, industry, academia, the 
media, and non-profits, to name a few. In my current role as a business 
development manager for Northrop Grumman Aerospace Systems, my academic 
background is fundamental to working with astronomers on missions in every 
stage. I find it very fulfilling to have a position that allows me to be 
part of the astronomy community, while also contributing through writing, 
public speaking, and translating technical and scientific information to 
the broader public. I've learned as the scientific, technical, and 
political environments evolve, so do the opportunities to contribute and 
create new paths.

\vskip 6pt
{\it AAS Newsletter, 150 (Jan/Feb 2010)}
 
\newpage
 
\centerline{\bf\large Science Communication as a Press Officer}
\vskip 6pt
\centerline{Christine Pulliam (\href{mailto:cpulliam@cfa.harvard.edu}{cpulliam@cfa.harvard.edu})}
\vskip 4pt

In today's world, more than ever, science communication is a crucial part 
of the overall scientific enterprise. This is especially true in the field 
of astronomy. Much astronomy research is taxpayer funded; the taxpayers 
deserve to know what they're getting in exchange for their hard-earned 
dollars.

Every astronomer can play a role in educating the public about who we are 
and what we do, whether through lecturing, writing articles in popular 
publications, or authoring books. But some of us choose to make science 
communication our full-time avocation.

While still in graduate school, I realized that the path of the research 
astronomer was not for me. I liked operating a large telescope and 
gathering data, but I didn't enjoy spending six months analyzing that data 
to eke out a speck of new knowledge. Too little fun, too much tedium. On 
the other hand, I enjoyed teaching, or more broadly, communicating 
scientific concepts and the latest discoveries.

After earning a master's degree in astronomy. I quickly landed a position 
at the Brookhaven National Laboratory on Long Island as a public 
information officer. My role was to share information about the lab with 
reporters, local residents, and the general public. I got on-the-job 
training in writing press releases, organizing events and meetings, 
guiding tours, and speaking to groups -- all skills that I use to this 
day.

After a few years, I started looking for a job more in line with my 
interests. Professional networking led me to the Harvard-Smithsonian 
Center for Astrophysics and an opening in their press office.

My primary duties involve writing press releases and web features to 
publicize discoveries and the scientists who make them. I often describe 
my role as ``translator'' -- speaking to astronomers in the specialized 
language of Str{\"o}mgren spheres and acoustic oscillations, and then 
reworking that information into something the layperson can understand. I 
also coordinate public outreach events, provide speaker training, and 
supervise a corps of telescope volunteers who conduct skywatching after 
our public lectures.

Our organization hosts about 300 Ph.D.\ researchers, so discoveries flow 
thick and fast. Yet not every journal paper will warrant a press release. 
Some researchers are surprised when the press office declines to issue a 
release on, e.g., magnetic polarization of protostar environments. They 
forget that while a finding may be scientifically valuable, it's not 
necessarily something that will ``wow'' the public.

The flip side of this challenge is that when a dramatic discovery does 
come along, I get to be one of the first people to hear about it. I've 
been privileged to publicize such milestones as the first ``weather map'' 
of an extrasolar planet and the ``super-sizing'' of our galaxy via a new 
mass calculation. (The latter was even featured on The Colbert Report!)

Lest you think that no research is involved, our press office pays careful 
attention to readership numbers for print and online media outlets. We 
also study audience interests and overall media trends. We have to 
constantly adapt our communications techniques to effectively reach our 
target audiences.

The media world is changing fast. Print is waning, while the web is ever 
growing. Twitter, YouTube, Facebook, and other sites offer a new way to 
reach people directly. Yet the audiences are becoming more fragmented, 
making it tough to expand beyond the science-interested to the more 
nebulous ``general public.''

The world of science communication is changing as well. Most newspapers no 
longer have a journalist dedicated to science topics. Instead, they rely 
on generalists or even crib directly from press releases. As a result, the 
role of press officer is taking on greater importance.

To be a good press officer, you must understand the science, but you must 
also be able to convey that understanding in simple language and sound 
bites. When a typical TV news story is three sentences long, you don't 
have any time to waste.

To prepare for a career in science communication, classes or seminars on 
journalism certainly help. Dedicated programs in science writing also 
exist at a handful of universities. Many of the required skills, though, 
are best picked up on the job.

The National Science Writers Association is the biggest clearinghouse for 
science writing jobs, including both salaried positions and internships. 
You can also check with specific organizations that interest you, such as 
publications ({\it Sky \& Telescope, Astronomy}) or NASA centers (Goddard, 
Spitzer, etc.).

In summary, if you enjoy telling people about amazing discoveries, 
consider becoming a press officer. You, too, can help to bring astronomy 
-- one of the few sciences that inspires an emotional connection with 
people -- to the world's attention.

\vskip 6pt
{\it AAS Newsletter, 149 (Nov/Dec 2009)}

\newpage
 
\centerline{\bf\large Jobs in Industry}
\vskip 6pt
\centerline{Peter Williams (\href{mailto:peter.todd.williams@gmail.com}{peter.todd.williams@gmail.com})}
\vskip 4pt

What do you expect from a job in industry? I made the leap. I expected a 
complete loss of autonomy, a stressful, degrading work environment, 
coworkers and bosses who wouldn't understand my background, but at least a 
good paycheck. In all regards but the last, I was wrong. I've had a blast 
ever since.

Academic life tends to paint the corporate world with the washed-out, 
uniform grays of a smoggy distant cityscape. The truth is that some 
corporations are everything you fear, but others are rewarding, exciting 
places to work. You can have a rich, challenging career in industry, and 
never look back. Or, you can regret making the move forever.

I began my search in industry when I couldn't land a job in academia. At 
first, I got turned down for every job I applied for -- even a gandy 
dancer. After many months, I eventually learned the arts of the resume (do 
not use a CV!), the cover letter, the presentation, the interview, and how 
not to appear desperate. That plus a good deal of luck, and I landed my 
first job in industry as a computational physicist.

Yes, I work in a cubicle. No, my job doesn't have the sex appeal of 
astrophysics. I haven't published in a while either. Frankly, I labor in 
relative obscurity. Where's the upside?

Let me begin with the autonomy I mentioned earlier: I have far more 
autonomy now than I ever did in astrophysics.

Autonomy is supposed to be one of the prime perks of academia. That is the 
myth, but at least in my world, reality fell far short. By the end of grad 
school I had developed a long-range vision for a program of research on 
jets, but I found I had no freedom to implement it. Each postdoc meant 
long hours devoted to somebody else's project; there was no way to do my 
own work. In industry, we would say that my interests and those of my 
bosses were fundamentally misaligned. I became stuck in the paradox that 
to do what was best for my career meant to throw my career in the toilet.

In contrast, in industry, I have complete control of my multi-year 
project. There is nobody breathing down my neck. There is no advisor 
poking his head in for frequent updates; there is no grant committee that 
needs a written progress update. In fact, there is no grant committee at 
all. If I want to do something, I do it. Succeed or fail, I own it.

And failure isn't necessarily bad. If half of your projects don't end in 
complete failure, you're probably not pushing the boundaries enough. A 
good fraction of the equipment you buy you may find you don't actually 
need. That is to be expected. Nobody is going to pester you about it.

In fact, one of the hardest things for new hires to learn is how to act 
with their newfound autonomy. We have a hard time getting newbies to spend 
enough money! They are used to having to grovel for second-hand equipment. 
We don't do that. Time is worth more than money. If we need something, we 
buy it. In industry, you will never want for the tools you need to do your 
job.

If this sounds like being treated like an adult, it's not a mistake. While 
there is a hierarchy, it's oddly liberating.

There is a hierarchy in academia as well, but we pretend there isn't. 
Again, the myth in academia is that all inputs are judged on their merits, 
not on who is making the input. The reality, I would argue, is that some 
people are not to be interrupted when making a point at a seminar, some 
people's theories must be attacked obliquely, and some people's grants get 
viewed in a more favorable light. Deference must be paid to the experts, 
even if they are wrong.

There is no such concept where I work. There are no sacred cows or 
powdered wigs. The hierarchy is explicit; everyone has a boss, but the 
hierarchy we have exists only in executive function, not expertise. I am 
the expert on my subject matter, and in that regard my bosses pay 
deference to me, not the other way around.

So what's my day like? Well, I spend most of my time doing actual physics. 
These aren't trivial problems; when I finish work on a topic it is worthy 
of publication if I so desire, but I'm usually off to something new. I 
have a primary project, which is physics code development. I also have 
various secondary problems I'm working on at any given time. Some of these 
are problems that others have posed to me; I don't have to solve those, 
but it's a point of pride that I do. Other projects are inventions I've 
thought up that sooner or later might end up being patented and put into a 
product that ships.

My colleagues are all competent and industrious; otherwise, they would be 
shown the door -- we don't have tenure. By no means are they researchers 
who ``couldn't hack it'' in academia. They are quite smart but slow to 
show it off; nobody cares how smart you are if you can't play well with 
others, and a brilliant person who nobody can stand to work with will soon 
be out of a job. The only weaknesses my coworkers show is a tendency to 
smile a lot, leave work at five, and not come in on the weekends. After 
decades in academia I for one have finally learned what regular sleep 
feels like.

I never doubt for a moment that my work isn't important or useful. I know 
that it leads to better chemical detection devices that help in some small 
way to make for a better world through the advances in medical technology 
that they enable. For many of my colleagues this is one of the prime 
motivators. When we die no textbook will eulogize our accomplishments; 
only in anonymity do we make small advances that lead to big improvements 
in society's health and safety, and that alone helps us to get out of bed 
in the morning.

You will never ever have job security in industry. That's a definite 
downside. But, as long as you are working, you should be paid well. Just 
like you, if I had wanted money to begin with, I wouldn't have gone into 
astrophysics. Quite frankly, if I had found autonomy in academia, I'd be 
happy as a clam earning a small fraction of my current pay.

But let me tell you, it's not bad being paid more than your professors. It 
may sound crass to discuss numbers but in my view it's essential because 
otherwise you'll get taken advantage of. With a physics Ph.D.\ and a good 
resume, in Silicon Valley, your starting salary as a minimum should begin 
with a ``1''. (Knock off 20\% for astronomy -- sorry!). Wall Street will 
pay several times that, even now. Close to half a million is not unheard 
of. Whatever you do, do NOT settle for something like \$70k, because I 
guarantee you that jobs that pay that level won't challenge you and won't 
offer upward mobility. This is not academia. Less is not more. Less is 
less.

\vskip 6pt
{\it AAS Newsletter, 147 (Jul/Aug 2009)}

\newpage
 
\centerline{\bf\large Back to School: A Ph.D.\ Enters the Classroom}
\vskip 6pt
\centerline{Joshua Roth (\href{mailto:jrothastro@yahoo.com}{jrothastro@yahoo.com})}
\vskip 4pt

``Isn't that a waste of your education?'' So spoke a few when I 
contemplated becoming a high-school physics teacher. The education in 
question: a bachelor's at U. C. Berkeley and an M. S. and a Ph.D.\ at 
Caltech (all in astronomy). But I can't honestly think of a better way to 
use my training and subsequent work experience.

Still, no one should enter the K-12 classroom out of a sense of 
obligation. My motivations were more modest: With two kids of my own, I 
wanted to see public schools from the inside. I wanted to see if my real 
or imagined gifts for communicating science -- honed during 11 years at 
{\it Sky \& Telescope Magazine} -- could be channeled into coherent lesson 
plans and engaging activities. Last but not least, I wanted to spend time 
in the company of people who are, for the most part, refreshingly honest 
and idealistic.

With rare exceptions, those terms describe my students and colleagues 
both. What's more, schoolteachers are, if possible, even more generous 
than research astronomers, freely sharing lesson plans or entire online 
courses that took them hundreds of hours to create. They are articulate, 
engaging, quirky, intellectually rigorous, and resourceful. And they are, 
for the most part, well respected in the communities they live in and 
serve (my opening quote notwithstanding). What a great crowd to rub elbows 
with at one's workplace!

That said, your fate as a new K-12 teacher depends very much on where you 
will put in that all-important first year. Some districts leave newbies to 
fend for themselves, while others generously mentor them. And demands on a 
K-12 teacher are intense even in the best-run, most resource-rich 
districts. If you aren't already close to someone who does this for a 
living, read {\it Teachers Have It Easy: The Big Sacrifices and Small 
Salaries of America's Teachers}. There you can read about educators who 
(like me) get out of bed at 5:30 a.m. and (like me) fall asleep at 10 p.m. 
beside piles of paperwork, making hundreds of high-stakes, on-the-fly 
decisions in the interim -- all under stringent legal and ethical 
constraints and the glare of a hypercritical public with outsized 
expectations.

Like research astronomy, K-12 teaching is a relatively illiquid labor 
market. Most jobs start with the new school year (though hiring may begin 
the previous spring). You generally need a teaching license to enter the 
classroom (though exceptions often are made for math and science teachers, 
especially in underserved districts; and some states grant temporary 
licenses to those capable of passing certain tests). And pay and job 
security are primarily based upon seniority (though charter schools and 
``independent,'' or private, schools, break this mold to varying degrees).

To earn the equivalent of tenure in public schools, you'll almost 
certainly have to take graduate-level education courses and possibly even 
earn another advanced degree -- all to earn about as much as a postdoc as 
long as you remain in the classroom (the real money is in administration 
or consulting). At least astronomers can take advantage of the fact that 
physics, chemistry, and math positions are among the hardest to fill 
(opportunities to actually teach astronomy, alas, are relatively rare, 
though you may be able to start up a club or independent-study program).

Many first teach as students in ``ed schools,'' under supervision. But I 
got my feet wet flying solo, covering for a physics teacher on maternity 
leave. Such ``long-term-subbing'' is a great low-stakes way to see what 
it's like being in front of a captive teenage audience without making a 
long-term commitment. Opportunities often appear on craigslist or 
school-district Web sites; but I think it best knock on doors and 
introduce yourself to superintendents, principals, and above all 
department heads (often called ``curriculum coordinators''). You'll almost 
certainly get an appointment -- folks with Ph.D.'s and published research 
don't walk into their offices every day!

I could say lots more about what has turned out to be the most challenging 
and the most rewarding work I've yet done, but I've already exceeded my 
ostensible 600-word limit. Any AAS member seriously contemplating a career 
in K-12 education can drop me an e-mail at \href{mailto:jrothastro@yahoo.com}{jrothastro@yahoo.com}; 
I'll reply with a list of my favorite teaching-related books and Web resources.

\vskip 6pt
{\it AAS Newsletter, 146 (May/Jun 2009)}
 
\newpage
 
\centerline{\bf\large Working at a Soft-Money Institute}
\vskip 6pt
\centerline{Barb Whitney (\href{mailto:bwhitney@spacescience.org}{bwhitney@spacescience.org})}
\vskip 4pt

Soft-money scientists are people who support themselves through research 
grants. This career path scares a lot of people because it is not secure 
employment. However, the amount of funding that a soft-money scientist 
brings in for their salary is often not much more than what faculty at 
research universities bring in for students, postdocs, and/or summer 
salary support. And there are many benefits to the soft-money lifestyle, 
as described below.

You can be a soft-money scientist at a variety of places: research 
universities, small colleges, observatories, government laboratories, and 
soft-money institutes. Each of these places will have a different culture, 
so my experiences won't apply to all of them. I am an off-site researcher 
at a small non-profit research and education organization (a soft-money 
institute). I work from my home in Wisconsin and my institute is in 
Colorado. So my boss is a comfortable 2000 miles away. The institute was 
created, and still has as its mission, to administer grants for scientists 
and educators. As long as I bring in enough grant money to support myself, 
I am an employee with full benefits, which include health insurance, life 
insurance, and retirement -- the same benefits I received when I was at a 
university. The administrators are friendly and easy to work with, and 
they do as much of the administrative work as possible -- leaving me with 
more time to concentrate on research. The overhead rate is reasonable, and 
allows me to use more of my funding for salary than at most traditional 
institutions. My salary and pay raises are determined by my performance 
and ``market'' values. Like most astronomers, I work with collaborators at 
a local university, and with several other scientists from around the 
world.

I've already outlined some of the positive features of this career path. 
Here are a few more: 1) You can live where you want. One of my colleagues 
lives on an island; another has moved to follow his wife's jobs, since she 
has a more lucrative career. 2) Working from home can be convenient for 
various family situations, like caring for small children or the elderly. 
3) Writing proposals helps you define and refine your research ideas and 
goals. 4) You can be flexible and creative about how you pursue your 
teaching/outreach goals, if you have any. For example, you can teach at a 
local community college or volunteer at museums. 5) Current computer and 
networking technologies make it easy to set up a fully functional office 
at your institute and/or home. 6) On a more personal note, the lack of job 
security gives me a different perspective than a lot of my colleagues on 
more traditional paths. It's good to have a Plan B for when the money runs 
out, and it's a worthwhile exercise thinking about it. I budget my future 
expenses assuming that I will only have half-time support, so I live a 
more modest lifestyle than I might otherwise. I have placed myself in a 
financial position that would allow me to take a lower-paying but more 
rewarding job if the opportunity arises. All of this makes me appreciative 
and grateful on a regular basis that I still get to do astronomy. 7) 
Working at home has also motivated me to become more involved in my local 
community, and my social life is consequently more varied and interesting 
than it was when I worked at a university.

Some drawbacks of being a soft-money scientist are: 1) For most people, 
the biggest drawback is the lack of job security. A scientist at a 
university might be able to move to other projects or find other jobs 
within the university if they run out of their own funding. 2) Small 
soft-money institutes don't have the facilities that large universities 
do, such as shares in large telescopes. I rely on national facilities. 3) 
Some people really don't like writing proposals -- in which case you 
probably don't want to be a professor at a research institution either. 4) 
Working off-site from a home office can be a difficult transition for some 
people. But only a minority of soft-money scientists work from home. 5) 
For me personally, I don't interact enough with the outstanding astronomy 
department in my hometown, and this is something I need to work on.

How do you become a soft money scientist? The same training as for a 
faculty job at a research institution works well for the soft-money 
scientist. The postdoctoral stage is a good training environment: you want 
to build your network of colleagues at one or two places besides your 
Ph.D.\ institution, attend meetings, publish papers, and learn how to 
write proposals. You want to develop a research program that is considered 
useful by your peers (and future review panels). You may not want to do 
exactly what everyone else is doing (jumping on the bandwagons) if you 
want to stand out a bit.  You can learn how to write good proposals by 
watching and doing. Collaborate with someone more senior and write 
proposals with them, both as PI and co-I. Volunteer to sit on review 
panels. Take advice to heart -- when someone tells you they don't get the 
point of your proposal, either you need to improve your message delivery 
or maybe your idea isn't so great after all. It really is a useful way to 
think through your ideas.

Once you've become a soft-money scientist, how do you thrive at it? Again 
this is similar to how faculty and other researchers thrive. I find that 
going to meetings or traveling to work with my collaborators usually 
provides a huge boost in enthusiasm for my projects and ideas. You need to 
publish a few papers a year, not necessarily as first author. Here it is 
difficult for me to offer too much advice since I make so many mistakes 
myself. For example, one thing to watch out for, which I do all the time, 
is taking on too many collaborations and not finding time to do your own 
research. Another thing that would be great to have, which I don't, is 
good time-management skills. However, I do follow my final piece of advice 
well, and I recommend it highly: Have fun with your research, and never 
stop learning!

\vskip 6pt
{\it AAS Newsletter, 143 (Nov/Dec 2008)}
 
\newpage
 
\centerline{\bf\large Balancing Research and Service at NOAO}
\vskip 6pt
\centerline{Knut Olsen (\href{mailto:kolsen@noao.edu}{kolsen@noao.edu})}
\vskip 4pt

When I left graduate school to take up a postdoc at the Cerro Tololo 
Inter-American Observatory in Chile (a division of NOAO), I had experience 
in research and in teaching, but very little with the inner workings of an 
observatory. I had been on a few observing runs, but never visited a major 
observatory, CTIO included. I was excited by the opportunity to learn 
about the technical aspects of observational astronomy -- to experience 
first-hand how the tools on which our work depends are developed and 
maintained, while at the same time getting my research career going. My 
first impression of CTIO was that the staff were extremely dedicated to 
the mission of the observatory -- providing first-rate facilities for the 
US astronomical community in the southern hemisphere -- but at the 
apparent expense of their own research. As my main ambition was still to 
do research, I at first felt a little out of place.

As I got to know CTIO better, however, I found that most of the staff 
were, after all, dedicated to their research. All felt strongly that 
having a scientific staff active in research was absolutely essential for 
the observatory. Observatory operations, user support, and future planning 
all require expert scientific input. NOAO felt that having staff motivated 
by their own scientific self-interest was the most natural way to provide 
excellent support. Although the observatory considered service and 
research to be roughly equally important, maintaining the balance between 
them could be tricky, as I learned when I was later hired as a CTIO staff 
astronomer. Our hallway discussions and meetings were invariably dominated 
by observatory matters, as these were often the most pressing, while 
personal research tended to be done in relative quiet, and could thus be 
disrupted; hence my initial impression that research was getting short 
shrift. Over time, though, I found that there are some ways to help 
maintain that balance, as I learned from others and from my own trial and 
error:

{\bf Align your service responsibilities with your research goals:} It may 
sound obvious, but making sure that your service overlaps with your 
research is key to making sure that you have enough time for research. A 
good and easy way to start is to make sure that the observatory's 
facilities feature prominently in your own research, and to offer to help 
support or be the instrument scientist for an instrument that you 
frequently use. This is good for both you and the observatory, as your own 
experience and needs can be used to help improve the experience for all 
users of the instrument.  At CTIO, I was Hydra instrument scientist and on 
the team that supported the Mosaic imager, both of which I was using in my 
research. I had a strong interest in working out the kinks with Hydra, 
because I knew how frustrating it was to have my night lost due to 
repeated instrument failures. My efforts paid off in that I got the data I 
needed from Hydra, and, because others came to see me as the Hydra 
``expert'', was able to join some very fun and productive collaborations 
that also needed Hydra.

If your research begins to outgrow the capabilities of the instruments 
that you support, then it's time to start thinking of ways to upgrade 
them, or consider new instruments (or facilities) that would do your 
science better. Working on science cases for facilities that remain in the 
distant future can be very helpful in guiding your research, as you will 
come up with very exciting ideas that will probably require significant 
groundwork to be done with current facilities. I am currently very excited 
by the ability of ground-based adaptive optics to study the high surface 
brightness bulges and disks of nearby galaxies, an interest which arose 
entirely because I contributed to the science case for a GSMT.

{\bf Talk to people:} Talking to others about their research is a good way 
to maintain focus on your own.  Make an effort to chat with users that you 
support about what they aim to do with their telescope time, and how your 
observatory's facilities fit in to their program. Offering to organize 
colloquia or seminars for a period is a good way to meet other 
astronomers, especially if your budget includes buying them lunch or 
dinner. In Chile, going out to eat with visiting astronomers was an 
especially good way to establish close relationships with them, since 
besides being an opportunity for you to talk to them about their research, 
it was a chance for them to get a local's perspective on Chilean culture. 
Serving on the observatory TAC is another good way to grow bonds with 
fellow astronomers, while also giving you an overview of a broad section 
of the current observational enterprise.

Working with students and/or postdocs is also very useful, although it 
means removing yourself a step from the actual work in exchange for the 
reward of helping someone else to learn and discover. At CTIO, we had 
three-month visits by undergraduate REU students every southern summer, 
and several graduate students and postdocs on extended visits. They were 
an integral part of the scientific culture, as they were the only people 
at the observatory fully dedicated to research.

{\bf Streamline your service work:} If parts of your service tasks become 
routine, they can sometimes be made more efficient, leaving you more time 
for research. For instance, for new users of Mosaic and Hydra, we would 
generally have a staff member present for part of the first night to help 
get the observers going, which involved 1.5 hours driving from La Serena 
each way. We eliminated the need to drive to the summit by installing 
videoconferencing equipment in the control room and downtown, and running 
the instrument user interface within a remote desktop environment. This 
way, the support astronomer downtown could be virtually present in the 
control room, having access to nearly all of the same screens available to 
the observer, while saving the large overhead of travel to the summit.

{\bf Avoid overcommitment:} This is easier said than done, of course, but 
bears mentioning. In particular, it is important to avoid taking on too 
many service responsibilities that have no overlap with your research. The 
truth is that every observatory has a number of tasks unrelated to 
anyone's research that need to be done by staff astronomers, {\it e.g.} 
editing newsletters, overseeing web page updates, maintaining 
documentation, and serving on internal committees.  Doing this work is 
important for the success of the observatory and thus should not be 
refused lightly if you are asked to help. However, this does not mean that 
you have to shoulder the burden indefinitely. If a task does not have an 
easily defined point of completion, then you might ask from the beginning 
to have a time limit on your involvement, after which your 
responsibilities are handed off to someone else.

{\bf Define your success by your research accomplishments:} You may find 
that many of your service responsibilities give you a lot of satisfaction. 
For instance, getting a broken instrument working again, discovering and 
explaining issues with data obtained from the observatory's facilities, 
helping users understand their strange data, or finishing off reports from 
committees on which you've served can all give you a strong sense of 
accomplishment. If you wish to remain scientifically active, however, it's 
important that you not become content with these accomplishments, but 
judge yourself mainly by the success of your research, which in many cases 
can be more difficult and tiresome to achieve. If you find yourself 
drawing much more satisfaction from your service work than your research, 
then you might consider reducing your research time in exchange for a 
larger service load. Indeed, observatories depend heavily on having a 
portion of their staff dedicated primarily to service for their success. 
If you envision yourself mainly as a research astronomer, however, don't 
be tempted to focus more on service, where you have achieved success, 
because you feel stuck in your research. Work through it.

In summary, working as an NOAO staff astronomer has been rewarding and 
exciting.  It is place to gain a good technical understanding of 
telescopes and instruments, be involved in providing access and developing 
cutting-edge facilities, and have the opportunity to contribute to 
initiatives of national importance to astronomy.  If you can maintain the 
balance between research and service, it's also a great place to have a 
productive scientific career.

\vskip 6pt
{\it AAS Newsletter, 142 (Sep/Oct 2008)}
 
\newpage
 
\centerline{\bf\large Succeeding in a Large Research Collaboration}
\vskip 6pt
\centerline{Andy Howell (\href{mailto:ahowell@lcogt.net}{ahowell@lcogt.net})}
\vskip 4pt

Large collaborations are now common in astronomy, and have produced some 
of the most transformative recent discoveries. Yet, when I finished grad 
school and was given the opportunity to work for a big team, few of my 
elders had any useful advice for how to succeed in such an environment. If 
they said anything, it was often a variation on, ``Don't join one.'' Fear 
of large collaborations is pervasive in astronomy, partly because it goes 
against our romantic notion of the lone astronomer on the mountaintop, but 
also because we rightly worry about being rendered anonymous in a sea of 
authors. But the benefits of large collaborations cannot be overlooked -- 
access to enormous data sets, learning from a wide array of people, and 
the possibility of making truly monumental discoveries. The trick is 
avoiding the traps. I had to learn the hard way -- by repeated failure!  
But after several collaborations, and much trial and error, I have 
identified a few strategies that worked.

{\bf Write papers:} In any large collaboration there are always fires to 
be put out. There are proposals and software to write, and a never-ending 
flow of data to reduce. There will be pressure to get the problem {\it du 
jour} solved, and it is the nature of peer pressure that the group wants 
you to put top priority on what is in the collective interest. But the 
only one who can act in your own interest is you. At the end of the day 
the most important metric for career success is your first-author papers, 
so you must be aggressive about writing them. I knew half a dozen selfless 
people, who always put the interests of the collaboration first, and fell 
under the illusion that success for the team equated to personal success 
for them. Now they are out of astronomy, because when their postdoctoral 
terms were finished, they had no first-author papers to show for it.

{\bf Have your collaboration duties match your scientific interests:} 
There is no getting around having to do the basic work required to keep 
the collaboration going. But this is much more enjoyable when your duties 
are necessary for your own paper, and everyone else gets to benefit as a 
side product. For example, in the Supernova Legacy Survey (SNLS), one 
person who discovers the supernovae is working on SN rates, one person 
responsible for spectroscopy uses it to produce papers on SN physics, and 
another member doing photometric calibration is writing one of the 
cosmology papers. Each person's work is dependent on everyone else's, but 
we are motivated by self-interest to do the best possible job in our area.

{\bf Encourage competition:} It may seem counterintuitive that a key to 
success is encouraging others to try to beat you to publication, but the 
only collaboration I have been in that encourages ruthless competition has 
had a 3-5 times higher publication rate than other similar collaborations. 
Five times! In the SNLS, no science is reserved -- it is all fair game for 
anyone to do at any time. Sure, this means that sometimes people's toes 
get stepped on. There have been several cases where two individuals were 
working on the same subject. Sometimes, if one person wasn't very far 
along they just dropped the research. Other times, either the analyses had 
to be merged, or two separate papers were written using different 
techniques, focusing on different parts of the problem. This is somewhat 
inefficient, though far outweighed by the overall efficiency gain. There 
were often heated discussions, and sometimes hurt feelings, but in every 
case the resulting science was stronger, and was produced at a much faster 
pace than in collaborations where science is pre-allocated. Besides, in my 
experience, personal disagreements happen at about the same rate no matter 
the organizational structure -- they seem to be more a function of 
individual personalities.

{\bf Think creatively:} I am amazed at the number of people in any 
collaboration who don't care to come up with original ideas for papers -- 
most simply want to repeat older work with better data. But large 
collaborations allow new kinds of studies that have never been done 
before, and these new, creative uses for the data often become the 
most-cited papers. The added benefit is that while everyone in the 
collaboration is fighting over the old ideas, you have exclusive access to 
your own fresh ones.

But creative ideas don't come easily. It is essential to be well read both 
inside and outside of your field. From the inside, there are new 
theoretical ideas to be tested, and papers from outside your field can 
suggest innovative techniques and new uses for your data. Going to talks 
and conferences, reading astro-ph, and having random conversations with 
colleagues over coffee, especially in areas outside of your specialty, are 
every bit as essential as hammering away on data.

{\bf Just do it:} The organization and politics of large collaborations 
can be complex, so sometimes asking permission to do something can take 
months to get a resolution. And it is often not in the interest of others 
to close off their options by giving you permission to do a certain study. 
Even if they are over-committed now, they may want to focus on that topic 
in the future. The largest time waster in any collaboration is arguing 
over hypothetical future outcomes, the majority of which never come to 
pass. Instead, if you just do the science (assuming it isn't too much of a 
transgression of collaboration rules), few will argue when there is a 
finished paper. Everyone benefits from a new publication, and it is hard 
for others to argue that you shouldn't have written a paper because they 
fantasized about writing a similar one. Instead the discussion shifts to 
the interesting new results.

In summary, collaborations can present unique barriers to paper writing -- 
group obligations, pre-allocated science, and the politics of the 
hypothetical. But if you can remove the obstacles, the papers will flow, 
and everyone stands to gain.

\vskip 6pt
{\it AAS Newsletter, 141 (Jul/Aug 2008)}
 
\newpage
 
\centerline{\bf\large Preparing for the College Teaching Job Market}
\vskip 6pt
\centerline{Luke Keller (\href{mailto:lkeller@ithaca.edu}{lkeller@ithaca.edu})}
\vskip 4pt

I once had a cross-country running coach who advised that if you want to 
run fast, you need to run fast. Having been on three faculty hiring 
committees over the past five years, I see that a similar rule goes for 
landing a teaching-oriented college faculty job: If you want to be a 
teacher you need to teach. Teaching experience is the single most 
important aspect of an application to be a professor at a college or 
university where undergraduate teaching is the primary focus. If at all 
possible it should be teaching as the instructor of record (you are in 
charge) teaching a college-level course in physics or astronomy. It's not 
enough for teaching to be important to you -- after all, teaching is 
important to most people looking for this kind of faculty position -- nor 
is it enough to have a well-articulated teaching philosophy, though this 
is helpful and often requested in applications. What you need to 
demonstrate clearly throughout your application packet is that you have 
designed and taught at least one course, written exams, interacted with 
students as their primary instructor/professor, and that you, your 
students, and your teaching colleagues think it would be good for you to 
do it again.

There are many paths to a teaching-focused faculty position. The following 
suggestions are intended to help current graduate students who have no 
previous teaching experience and who want to apply for college-level 
teaching jobs immediately after graduating or after a short postdoc 
career:

\begin{itemize}
\item Getting teaching experience that will convince a hiring committee to 
invite you for an interview is the most important step. If you are lucky 
enough to be at a school that will let you teach your own class, take 
advantage of your good fortune. If not, look at the on-line job listings 
at a nearby small college or community college. Contact the physics 
department chair with your willingness and interest in teaching, even if 
there are no positions currently available. Most colleges have regular 
openings for teaching introductory physics and astronomy, and many are 
actually happy to help proto-professors develop and hone their teaching 
skills.

\item Offer to teach the lecture, not just a lab, since you most likely 
already have or will have lab instruction and tutoring experience as a TA. 
The goal is to get your own class and your name at the top of a syllabus 
that you have written. Without this kind of teaching experience, your 
chances of success on the job market are reduced.

\item More than one or two semesters of teaching while in grad school is 
not necessary. Keep in mind that you also need a well-established research 
program to land that college job and for that you need to finish your 
dissertation! Most search committees are reluctant to hire a person into a 
tenure-track job who will not have defended by the time classes start. 
Teaching is fun and it gets easier with experience so you may be tempted 
to spend more time on teaching at the expense of your dissertation; avoid 
that trap.

\item Give your students the opportunity to evaluate your teaching even if 
the place you teach does not normally do student evaluations for adjunct 
faculty. This will give you feedback and material for your own letters of 
application and for your reference letter writers.

\item Identify at least one faculty member, either at the college where 
you teach or at your home institution, who will visit your class a few 
times over a semester or quarter and then write a letter of recommendation 
that emphasizes your teaching. Give this person access to your student 
evaluations. Ideally this will NOT be your dissertation advisor since s/he 
will most likely be writing you a research-oriented letter that will be 
read along side the teaching letter. Most teaching-oriented jobs will ask 
for at least one letter that focuses on your teaching experience and 
effectiveness. You should also ask your supervisor (where you teach) to 
visit your classes and be prepared to write a letter on your behalf.

\item It is rarely a good idea to try to teach -- especially your first 
course -- while taking a full load of graduate courses. Wait until you 
have completed your course work. When you do finish with classes, get that 
M.A. or M.S. Though rarely required to go on for the Ph.D.\ in astronomy 
or physics, a master's degree will help you get that first teaching job 
while still in grad school. Some jobs require the master's.

\item Get help and advice on teaching methods, interacting with students, 
etc., especially if your teaching job is not at your Ph.D.\ institution. 
The best way is to talk with your colleagues, especially those who are 
likely to write letters for you later. You should also plan to attend at 
least one teaching workshop or conference. Many professional conferences 
(including AAS meetings) showcase pedagogy research and have optional 
teaching workshops for college-level teachers of physics and astronomy. 
The American Association of Physics Teachers is another excellent 
resource.

\item Balancing teaching with finishing your dissertation is essential and 
it's very much like balancing teaching and research after you land that 
faculty job. Expect your first teaching experience to take a lot of time. 
In the balance, though, realize that teaching is a project that takes as 
much time as you give it so set boundaries. Remember to factor commuting 
time into your scheduling and planning and make sure that your 
dissertation advisor knows when not to expect you to be in your office or 
in the lab.

\item You may experience well-meaning resistance to your teaching project 
from your own teachers and advisors. This may be in the form of advice 
that concentrating on your research is the best way to prepare for the job 
market. The question you should ask is ``Which job market?'' Try to make 
it clear that you want a job that requires teaching experience and ask 
their advice about balancing teaching with research.
\end{itemize}

{\bf So you found the perfect job ad...}\\
Faculty hiring committees usually know exactly what they are looking for. 
Their teaching job must be `Plan A' for you. You are wasting your time and 
theirs if you just change the institution name at the top of your 
research/postdoc job applications. It is astonishing how many application 
letters I've seen that basically say, ``Look at my fabulous research and, 
by the way, I'm committed to excellence in teaching.'' Teaching must come 
first; after all it will have to come first if you get the job. Teaching 
must also be the primary emphasis of your CV; list your teaching and 
teaching-related experience first, then your research experience and 
publications. Your application letter should reflect the ad and use 
similar language to argue that you fit the position and have done your 
homework to learn about the department and program. If you really want the 
job, contact the hiring committee chairperson for more information so you 
can really tailor your application to the position.

{\bf Whether to postdoc...}\\
The above guidelines go for postdocs as well. Even if you have decided to 
concentrate for now on research, if you know you eventually want a 
teaching job you need to prove it to yourself and future hiring committees 
by teaching sooner rather than later. It is very hard to convince a 
committee of teachers that teaching is your number one priority when your 
CV has a long list of papers, but no teaching experience beyond a few TA 
jobs in grad school. Consider applying for one of the growing number of 
teaching postdocs that carry a small teaching load, sometimes just one 
class per year.

Most teaching-oriented colleges like to see postdoctoral research 
experience in addition to teaching experience. They would like to see 
evidence of a well-established, even if modest, research program that is 
obviously doable in parallel with a heavy teaching load and with no 
graduate students.

{\bf But I want to be a professor at a large research university...}\\
The tier-one research universities are placing increasing emphasis on 
undergraduate teaching skill, especially in their new faculty hires. Top 
that fabulous CV off with a semester at the front of a classroom.

The bottom line: if want to be a teacher, you need to teach.

\vskip 6pt
{\it AAS Newsletter, 140 (May/Jun 2008)}

\end{document}